# The Magic XRoom: A Flexible VR Platform for Controlled Emotion Elicitation and Recognition

S. M. Hossein Mousavi[*]
Matteo Besenzoni[*]
Davide Andreoletti[*]
Achille Peternier[*]
Silvia Giordano[*]

University of Applied Sciences and Arts of Southern Switzerland, Lugano, Switzerland

*seyed.mousavi@supsi.ch, matteo.besenzoni@supsi.ch, davide.andreoletti@supsi.ch, achille.peternier@supsi.ch, silvia.giordano@supsi.ch*

**Abstract**: Affective computing has recently gained popularity, especially in the field of human-computer interaction systems, where effectively evoking and detecting emotions is of paramount importance to enhance users' experience. However, several issues are hindering progress in the field. In fact, the complexity of emotions makes it difficult to understand their triggers and control their elicitation. Additionally, effective emotion recognition requires analyzing multiple sensor data, such as facial expressions and physiological signals. These factors combined make it hard to collect high-quality datasets that can be used for research purposes (e.g., development of emotion recognition algorithms). Despite these challenges, Virtual Reality (VR) holds promise as a solution. By providing a controlled and immersive environment, VR enables the replication of real-world emotional experiences and facilitates the tracking of signals indicative of emotional states. However, controlling emotion elicitation remains a challenging task also within VR. This research paper introduces the Magic Xroom, a VR platform designed to enhance control over emotion elicitation by leveraging the theory of flow. This theory establishes a mapping between an individual's skill levels, task difficulty, and perceived emotions. In the Magic Xroom, the user's skill level is continuously assessed, and task difficulty is adjusted accordingly to evoke specific emotions. Furthermore, user signals are collected using sensors, and virtual panels are utilized to determine the ground truth emotional states, making the Magic Xroom an ideal platform for collecting extensive datasets. The paper provides detailed implementation information, highlights the main properties of the Magic Xroom, and presents examples of virtual scenarios to illustrate its abilities and capabilities.

## 1 INTRODUCTION

Affective computing is an interdisciplinary field at the intersection of computer science, psychology, and cognitive science that primarily concentrates on the elicitation and recognition of emotions [1]. The task of emotion elicitation is aimed to evoke emotional responses in individuals, while emotion recognition focuses on the development of systems that can accurately detect human emotions from a set of observations (such as facial expressions). These areas encounter notable challenges, such as the precise control of elicitation factors to evoke specific target emotions and the accurate detection of emotions.

Virtual Reality (VR) [2] has emerged as a powerful tool to address these challenges in affective computing due to its unique properties. Firstly, its immersive and interactive nature enhances the potential for evoking stronger emotional responses in users. More specifically, VR creates realistic virtual environments which allow individuals to deeply engage with the environment and hence, elicit more intense emotions. Secondly, VR enables precise control over environmental factors, such as colors and sounds, allowing one to tailor the virtual experience to elicit specific emotions. As shown in Refs. [3, 4], this level of control enhances the accuracy and consistency of emotion elicitation. Lastly, VR offers the capability to track users' generated signals, including their movements through the VR headset and physiological signals captured by external devices, that can be used to develop systems for automatic emotion recognition. We note that, since VR elicits stronger and more spontaneous emotions, and tracking does not interfere with the user's experience, VR represents an ideal means for comprehensive data collection.

While VR offers numerous benefits, the process of controlling emotion elicitation remains a challenging task, as it is not yet clear which factors truly impact the evoked emotions. In fact, there is currently a lack of comprehensive guidelines based on robust theoretical foundations for conducting experiments in the field of emotion elicitation. In this paper, we claim that the theory of flow [5] can help increase the control over the emotion elicitation process. The theory of flow identifies a direct relationship between an individual's skill level, task difficulty, and emotional state. According to this theory, when an individual is fully immersed and engaged in an activity that matches her skill level, she experiences a state of flow associated with increased enjoyment and satisfaction. Conversely, when skills are insufficient for the task's difficulty, anxiety arises, and when skills surpass the task's difficulty, boredom ensues. We argue that, by accurately assessing a user's skill level and modifying the task difficulty accordingly, the desired emotions can be effectively elicited.

This paper describes our implementation of a VR application, referred to as the Magic Xroom, for the systematic experimentation with hypotheses on the main factors of emotion elicitation. In particular, we leverage the properties of VR, namely, its flexibility and ability to evoke strong emotions, and of the theory of flow, namely the connection between skills, difficulty, and emotion, to build a platform for controlled emotion elicitation, which will serve the purposes of collecting extensive datasets for further research in the field of affective computing. In the Magic Xroom, the skills of participants are continuously monitored to determine their baseline, and the difficulty of the tasks they perform within the VR environment is adapted accordingly, aiming to strike the ideal balance between challenge and skill level to induce the desired emotional state. By allowing for dynamic adjustments of task parameters, the Magic Xroom ensures a precise and controlled approach to emotion elicitation, which can be further improved by customizing contextual factors, such as lighting, sounds, and other stimuli. Moreover, to gain deeper insights into users' emotional states, the Magic Xroom equips users with a comprehensive sensor suite that monitors physiological responses including heart rate, skin conductance, and facial expressions. Please note that the measures collected by the sensors can always be synchronized, ensuring that the information from different sensors can be effectively combined. This approach offers increased flexibility within the system, as it does not mandate the simultaneous activation of all sensors during data collection. By allowing individual sensors to function independently, the Magic Xroom becomes a modular and adaptive system, accommodating various sensor configurations beyond the basic VR setup. Additionally, the platform incorporates virtual panels for users to provide explicit feedback, enabling the collection of ground truth emotional labels.

The remainder of the paper is structured as follows. Section 2 reviews existing works in the field of emotion elicitation, specifically focusing on VR-based solutions. Section 3 elaborates on the use of VR and the theory of flow for enhanced emotion elicitation. Section 4 focuses on the description of the Magic Xroom. Finally, Section 5 concludes the paper.

## 2 RELATED WORK

The majority of the literature on elicitation strategies primarily focuses on traditional media formats such as images [6], [7], [8], sound [9, 10], [11], and video [10, 12], [13]. These traditional media-based methods offer significant advantages in terms of scalability (as capturing and sharing pictures, audio, and video with a large user base is relatively straightforward) and flexibility (as it is easy to adjust key parameters, such as colors in images). However, one notable drawback of these methods is their limited immersion and engagement, which can impact the authenticity of emotional responses, which are not as strong or genuine as they would in real-life situations. To address this limitation, recent research has proposed using VR as an elicitation medium. VR-based methods offer immersive and multi-dimensional experiences [14], which significantly enhance users' emotional responses. For instance, Ref [15] demonstrated that immersive VR environments effectively enhance the emotional experience through visual stimuli, resulting in increased arousal and valence of the evoked emotions. Along this line, VR has been proven effective in eliciting the target emotion in training [16] and gaming activities [17, 18]. Then, several studies have considered the theory of flow in VR. For example, Ref. [16] investigated the factors that contribute to the experience of flow during virtual surfing, while Ref. [19] examined the relationship between flow state and learning outcomes in car detailing training. The findings of these studies indicated that individuals who experienced a flow state had superior learning outcomes. However, these works do not employ the theory of flow as a means for improving the control over emotion elicitation, as we instead do in the development of the Magic Xroom.

## 3  VIRTUAL REALITY AND THEORY OF FLOW FOR EMOTION ELICITATION

VR is a groundbreaking technology that merges the real and virtual worlds, immersing users in computer-generated 3-Dimensional (3-D) environments [20]. This immersive experience is made possible through specialized hardware such as Head Mounted Displays (HMDs) [21] and handheld controllers, enabling interactive and engaging encounters that surpass traditional screen-based interactions.

When it comes to emotion elicitation, VR holds several advantages over conventional media. By immersing users in dynamic virtual environments, VR has the potential to evoke more authentic and intense emotional responses. Additionally, VR offers a high level of control over environmental factors, allowing for the exploration of various conditions and hypotheses related to emotion elicitation. However, the absence of theoretical guidelines on effectively inducing target emotions has impeded progress in the field.

In this study, we propose integrating the theory of flow into the process of emotion elicitation. The theory of flow is a psychological theory developed by Mihaly Csikszentmihalyi that explores the relationship between task difficulty, individual skills, and emotional states. It is named after the concept of flow state, which occurs when the level of task difficulty aligns with an individual's skill level. Flow state is characterized by complete engagement, intense focus, a sense of timelessness, and a deep sense of satisfaction and fulfillment. Conversely, individuals with low skills facing highly challenging tasks often experience anxiety, while those with high skills performing easy tasks tend to feel bored.

By considering the connection between a user's skills, task difficulty, and emotions, we can adopt a more controlled approach to eliciting emotions. Manipulating the factors that trigger specific emotions allows us to fine-tune the elicitation process accordingly. Our primary objective is to manipulate environmental elements to induce targeted emotions, thereby enhancing the overall effectiveness of the emotion elicitation process. For instance, increasing task difficulty can elicit negative emotions, while decreasing difficulty can elicit positive emotions. To the best of our knowledge, this research represents the first attempt to leverage the theory of flow to achieve a more controlled and precise approach to emotion elicitation. In the next Section, we will describe the current implementation of the VR application that we realized for controlled emotion elicitation.

## 4    THE MAGIC XROOM

The Magic Xroom is an innovative and immersive standalone VR application developed using the Unity game engine[1] and the SteamVR framework[2]. The purpose of creating the Magic Xroom is to improve the process of evoking and identifying emotions in a controlled setting. Essentially, the Magic Xroom provides a controlled and consistent environment for conducting experiments and collecting data. This enables researchers to methodically investigate different facets of emotional reactions. As a result, it serves as a valuable resource for advancing research, creating more accurate emotion recognition algorithms, and refining emotional models. The subsequent sections outline the phases of emotion elicitation and emotional recognition in greater detail.

Emotion Elicitation. The Magic Xroom engages users through sensory inputs and interactive tasks to elicit emotional responses. Following the theory of flow, task parameters (e.g., difficulty levels) need to be dynamically adjusted based on users' skills. To reach this aim, users' skills are continuously monitored throughout the experience, enabling the Magic Xroom to modify the virtual environment based on their performance. Moreover, task difficulty is carefully calibrated to elicit the intended emotions while parameters such as the available experiment time can be adjusted to create situations that induce emotions such as anxiety or stress.

Emotion Recognition. To accurately capture the emotions experienced by users, the Magic Xroom equips individuals with a comprehensive set of sensors that continuously collect diverse data throughout their interaction with the application. These sensors monitor physiological responses, such as heart rate, skin conductance, and facial expressions, providing valuable insights into users' emotional states. More specifically, Galvanic Skin Response (GSR) units are used to capture the electrical characteristics (e.g., conductance) of the skin, and the Optical Pulse/PhotoPlethysmoGram (PPG) signal, which is then used to estimate heart rate variability. Additionally, VR eye trackers are used to measure eye movement and focus, while face and mouth trackers are used to capture mouth, jaw, and chin movements. Furthermore, explicit user feedback is collected through intuitive virtual panels, allowing users to provide ground truth emotional labels.

---

[1] https://unity.com/
[2] https://www.steamvr.com/

## 4.2 Key Advantages

The advantages of the Magic XRoom are the following:

- Controlled Emotion Elicitation: The Magic XRoom enables the manipulation of task parameters to induce targeted emotional responses, facilitating accurate analysis and evaluation of emotional states.
- Immersive Experiences: The Magic XRoom utilizes interactive tasks, audio cues, and realistic visuals to enhance emotional impact and elicit authentic responses.
- Scalability: With minimal hardware requirements, such as a VR headset and controllers, the Magic XRoom is a scalable tool for widespread utilization of emotion elicitation techniques through VR.
- Continual Advancements and Integration: The Magic XRoom seamlessly integrates novel experiences, technologies, and devices, ensuring compatibility with advancements in VR hardware and software. It currently supports virtual reality headsets and handheld controllers. Eye trackers, face trackers, and Shimmer GSR compatibility are currently being developed.
- Easiness of Data Collection: The Magic XRoom incorporates a comprehensive set of sensors to collect diverse data, enabling detailed analysis and evaluation of emotional responses.

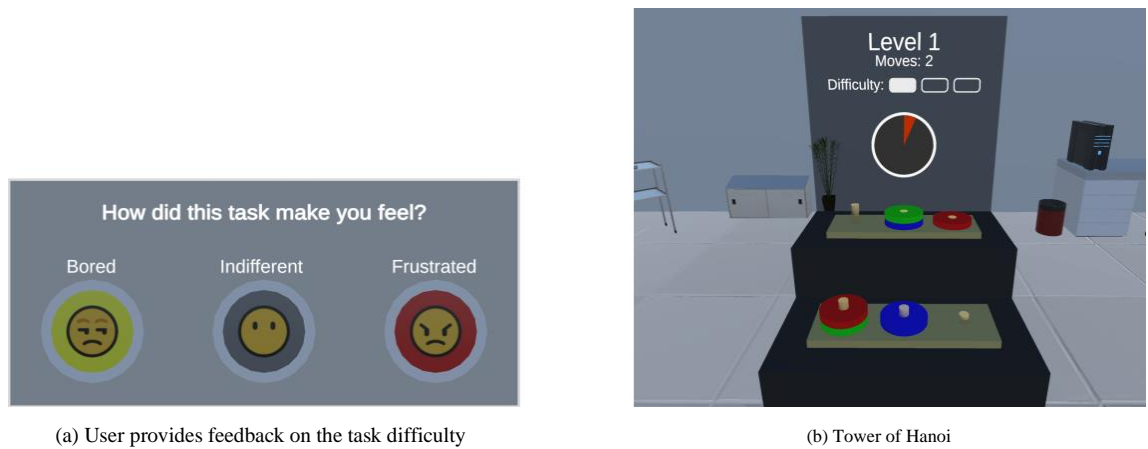

(a) User provides feedback on the task difficulty  (b) Tower of Hanoi

Fig. 1. Feedback and Tower of Hanoi

## 4.2 Examples Demonstration

A variation of the puzzle Tower of Hanoi has been created following this methodology 4.
The game consists of three rods and three disks, which can slide onto any rod 1b. The goal of the task is to replicate the configuration at the top with the disks and rods at the bottom with limited moves and time. A ticking sound is played to induce a sense of urgency, becoming faster in the last few seconds. The user controls the disks with the VR controllers and moves them one at a time from one rod to another one.

Tasks progress with increased difficulty, either by increasing the complexity of the target configuration or by reducing the available time and/or moves. The goal is to induce anxiety in the user due to the limited time available to think of a solution paired with limited movements, leading to one or two maximum sets of possible moves to solve the puzzle.

## 4.3 Future Work

The Magic Xroom is currently being implemented and there are several future tasks that are being planned. Firstly, a system that can automatically assess users' skills will be realized. This system will be crucial to adjust the difficulty level to evoke a specific emotion based on the theory of flow. One possible approach to evaluate and measure individuals' abilities automatically is to establish a set of task-dependent performance metrics. Secondly, we will develop additional virtual environments for the Magic Xroom. The tower of Hanoi was in fact merely a basic example of a virtual scene created. Lastly, the Magic Xroom will be utilized for extensive data collection. The collected data will inherently consist of multiple modalities as it will leverage various sensors that users will be equipped with (e.g., sensors embedded into the VR headset, along with additional devices, such as smartwatches for monitoring physiological parameters).

## 5 CONCLUSION

This paper outlines the current implementation of the Magic XRoom, an immersive virtual reality (VR) application designed to enhance control over the elicitation of emotions by applying the theory of flow. The paper discusses the main features of the Magic XRoom, highlights its key benefits, and provides an example of a virtual scenario that has already been developed, i.e., the Tower of Hanoi. The development of the platform is still in progress, with future efforts focusing primarily on incorporating essential functionalities, such as an automated assessment system for individuals' skills and the creation of additional virtual scenarios. Once the development phase is completed, the Magic XRoom will be utilized extensively for data collection purposes.

**ACKNOWLEDGMENTS**

A big thank you to Alessandro Ferrari and Denis Beqiraj for helping in the development of the Magic XRoom.